\documentclass{elektr}
\usepackage{hyperref}
\hypersetup{
colorlinks=true,
urlcolor=blue,
citecolor=blue}
\usepackage[all]{xy,xypic}
\usepackage{amsfonts,amssymb,amsmath,amsgen,amsopn,amsbsy,theorem,graphicx,epsfig}
\usepackage{eufrak,amscd,bezier,latexsym,mathrsfs,enumerate}\usepackage[utf8]{inputenc}\usepackage[english]{babel}
\usepackage[dvipsnames]{xcolor}
\usepackage[pagewise]{lineno}

\year{}
\vol{}
\fpage{}
\lpage{}
\doi{}

\title{Analysis of performance criteria for optimization based bearing only target tracking algorithms}

\author[SÖNMEZ AND HOCAOĞLU]{
\textbf{Hasan Hüseyin SÖNMEZ$^{1}$\thanks{h.sonmez@gtu.edu.tr}~, Ali Köksal HOCAOĞLU$^{1}$}\\
$^{1}$Electronics Engineering Dept., Gebze Technical University, Kocaeli, Turkey\\
\\ [1.8em]

\rec{.201}
\acc{.201}
\finv{..201}
}

\def\E{\ifmmode{\mathbb E}\else{$\mathbb E$}\fi} 
\def\N{\ifmmode{\mathbb N}\else{$\mathbb N$}\fi} 
\def\R{\ifmmode{\mathbb R}\else{$\mathbb R$}\fi} 
\def\Q{\ifmmode{\mathbb Q}\else{$\mathbb Q$}\fi} 
\def\C{\ifmmode{\mathbb C}\else{$\mathbb C$}\fi} 
\def\H{\ifmmode{\mathbb H}\else{$\mathbb H$}\fi} 
\def\Z{\ifmmode{\mathbb Z}\else{$\mathbb Z$}\fi} 
\def\P{\ifmmode{\mathbb P}\else{$\mathbb P$}\fi} 
\def\T{\ifmmode{\mathbb T}\else{$\mathbb T$}\fi} 
\def\SS{\ifmmode{\mathbb S}\else{$\mathbb S$}\fi} 
\def\DD{\ifmmode{\mathbb D}\else{$\mathbb D$}\fi} 

\newcommand{\bse}{\begin{subequations}}
\newcommand{\ese}{\end{subequations}}
\newcommand{\ben}{\begin{enumerate}}
\newcommand{\een}{\end{enumerate}}
\newcommand{\bens}{\begin{enumerate*}}
\newcommand{\eens}{\end{enumerate*}}
\newcommand{\be}{\begin{equation}}
\newcommand{\ee}{\end{equation}}
\newcommand{\bea}{\begin{eqnarray}}
\newcommand{\eea}{\end{eqnarray}}
\newcommand{\baa}{\begin{eqnarray*}}
\newcommand{\eaa}{\end{eqnarray*}}
\newcommand{\bc}{\begin{center}}
\newcommand{\ec}{\end{center}}

\newcommand{\vs}{\vspace}

\theoremstyle{corollary}

\theoremstyle{lemma}

\theoremstyle{proposition}

\theoremstyle{axiom}

\theoremstyle{conjecture}

\theoremstyle{example}

\theoremstyle{definition}

\theoremstyle{remark}

\begin{document}

\maketitle

\begin{abstract}
Target tracking problem has many practical applications in real life. In submarines, target tracking is done using, preferably, passive sensors. These sensors measure only the bearing angles between the observed target and the ownship. Therefore, this problem is generally referred as bearing only target tracking or target motion analysis. The classical approach is to use a state observer based filter, i.e. Extended Kalman Filter, to estimate the range, course and speed of the target, using only the bearings. In recent studies, the problem is solved as a global optimization problem by utilizing evolutionary algorithms with respect to some objective functions. In this study, we investigate the effect of the commonly used cost functions on the performance of the TMA algorithms. Particularly, we investigate the cost functions based on bearing differences and equidistant line segments. The simulation results show that the latter gives a sub-optimal solution to the target motion analysis problem, compared to the former.

\end{abstract}

\section{Introduction}
\label{Int}

In military underwater applications, such as navigation and tactical surveillance, the aim is to ensure the acoustic silence to remain undetected by the enemy (target) while tracking them. Therefore, in submarines, target tracking is mostly done using passive sonar sensors, instead of active sensors such as electromagnetic radars. Hence, in an acoustic environment, detecting the direction of a sound source by processing the information obtained with passive sensors and converting them to the bearing angles is of great importance \cite{sonmez_new_2017}. Obtained bearing angles are used by the analysis algorithm to estimate the next bearing, range, course, and speed of the target. This process is called Bearing Only Target Tracking (BOTT) or Target Motion Analysis (TMA).

TMA problem has been studied by many researchers since the World War 2. First studies prior to recent methods are proposed in the 80s by Aidala using Kalman filters both in cartesian and polar coordinates \cite{aidala_utilization_1973,aidala_utilization_1983}. Since TMA is a nonlinear state estimation problem, the measurement model violates the linearity assumption of the Kalman filters. Therefore, later studies are focused on the nonlinear and adaptive state observers and reported successful results.

The difficulty of the TMA problem mostly comes from the nonlinearity of the measurements and observability requirements. Even without noise, the nonlinearity may cause the state vector to be unobservable \cite{nardone_observability_1981}. In \cite{nardone_observability_1981}, Nardone and Aidala discussed the necessary and sufficient conditions for TMA observability issue. In \cite{song_observability_1996}, Song set up these observability conditions from a practical point of view. In theory, to obtain a unique solution, for a target which has \textit{N}-th order target dynamics, the observer must have at least (\textit{N+1})-th order dynamics \cite{song_observability_1996}. For example, for a target moving in a straight line with constant speed, the observer must have an accelerating motion, which is the common scenario for most of the work in literature. In \cite{le_cadre_discrete-time_1997}, Cadre and Jauffret presented an analysis of observability and estimability for TMA.

For the case non-maneuvering targets, most of the work in the literature based on state observer estimators. In \cite{spingarn_passive_1987}, discussed Extended Kalman Filter (EKF) and iterated EKF for the TMA problem. Since EKF has stabilization issues due to the linearized approximations, Aidala and Hammel proposed the modified polar (MP) coordinates in \cite{aidala_utilization_1983}, in order to stabilize EKF. In \cite{xu_single_2004}, Xu and Liping proposed Unscented Kalman Filter (UKF) which uses unscented transform to approximate the moments, instead of linearizing. They showed that UKF performs better accuracy and is more stable than EKF. In \cite{arulampalam_bearings-only_2004} Arulampalam et al. proposed using Particle Filters (PF) in a multiple model framework for maneuvering targets in both single sensor and multi-sensor cases. They reported that the PF based algorithms have a superior performance against the conventional IMM-based algorithms as in \cite{kirubarajan_bearings-only_2001}. Castaneda et al. proposed a recursive algorithm, based on PF, for ground moving targets from multiple bearings-only measurements in \cite{castaneda_new_2007}. The proposed algorithm also deals with target maneuvers. In \cite{miller_underwater_2018}, Miller and Miller proposed a pseudomeasurement method to linearize the problem.

In more recent works, Yu et al. \cite{yu_distributed_2016} proposed a distributed PF on spherical surfaces, which accounts for the curvature of the surface in the measurement model. They show that the distributed approach has an accuracy comparable to that of a centralized filter. In \cite{modalavalasa_new_2015}, Modalavalasa et al. extended the EKF algorithm to the bearing and elevation only tracking. Although the performance of this algorithm is slightly better than EKF, the convergence speed of the filter is low. In \cite{liang-qun_bearings-only_2015}, Liang-qun et al. proposed a BOTT algorithm for maneuvering targets based on truncated quadrature Kalman filter (TQKF). In the proposed method, prior distribution is modified based on the current measurement to reduce the effect of target maneuvers on the performance of the filter. It is shown that, by Monte Carlo simulations, the performance of the proposed algorithm is more accurate than UKF, quadrature KF, IMM-EKF and multiple-model Rao-Blackwellized PF \cite{Li2009197}. Leong et al. \cite{leong_gaussian-sum_2014} proposed a Gaussian-sum Cubature Kalman filter as a nonlinear filter. In \cite{nguyen_improved_2017}, Nguyen and Doğançay proposed a bias compansated pseudolinear Kalman filter and reported that mean squared error is close to the posterior Cramer-Rao lower bound at moderate noise levels. In \cite{zhang_piecewise_2017} et al. studied the observer trajectory optimization for BOTT. Zhang and Song \cite{zhang_improved_2017} proposed a novel iterated Gaussian mixture measurements filter to represent the measurement likelihood function and found it superior to existing methods.

Recently, random finite set (RFS) theory has been employed for the multi-target multi-sensor tracking systems and has been shown to result in computationally feasible multi-target tracking filters. In \cite{beard_bayesian_2015}, Beard et al. proposed a Bayesian filter, based on RFS, to track an unknown and varying number of targets for the multi-target bearing-only scenario. Xie and Song \cite{xie_improved_2017} proposed a Labeled Multi-Bernoulli (LMB) filter approach with reduced estimation error and tractable computational cost.

In addition to the state observer based methods, there are some works that define the TMA as an optimization problem. Because the TMA is a nonlinear problem, it can be seen as a multimodal optimization problem and global optimization algorithms are needed to find the optimal solution with respect to an objective function. The multimodal optimization problems are particularly hard to solve with gradient-based methods because they are usually stuck in local optima. For this reason, evolutionary algorithms often employed.

For the TMA based on evolutionary algorithms, the solution is usually represented with the initial range, course, and speed of the target, assuming constant course and speed for the target. In \cite{genc_bearing-only_2008}, Genç and Hocaoğlu proposed a nature-inspired evolutionary algorithm named Big Bang - Big Crunch (BB-BC). The algorithm generates random points in a search space in the Big Bang phase and shrinks to a single point which has the minimum cost in the Big Crunch phase. The paper assumes constant course and speed scenario, which is common for typical approaches for TMA. In \cite{genc_new_2010}, Genç proposed a new evolutionary approach, based on BB-BC and compared it with the original BB-BC algorithm and a simple GA. Another method is presented by İnce et al. in \cite{ince_evolutionary_2009}, using genetic algorithms (GA) and Monte Carlo simulations. The algorithm starts with an initial population and iteratively narrows the search space through Monte Carlo runs. Although it is reported for the algorithm to have a very good accuracy under different degrees of noise and observability conditions, it is computationally exhaustive. In \cite{sonmez_new_2017}, Sonmez et al. proposed to use an evolution strategies (ES) method called covariance matrix adaptation (CMA-ES). CMA-ES is proposed by Hansen and Ostermeier in \cite{hansen2001ecj} and it is shown that it is a reliable and competitive evolutionary algorithm, particularly for continuous global optimization problems \cite{hansen2004ecm,hansen_cma_2016}. In \cite{sonmez_new_2017}, it is shown that CMA-ES is capable of finding a solution for the TMA problem, under different degrees of noise, with high accuracy. Another example of evolutionary algorithm approach for the TMA problem is presented by Tokta et al. in \cite{tokta_target_2017}. It is a fitness adaptive version of the BB-BC algorithm.

In the case of evolutionary algorithms, defining a propoer cost function is crucial. For the studies at hand, there are two different cost functions defined for the TMA problem. First one is based on the difference of the bearing angles, which are obtained as observations and calculated using the candidate solution \cite{ince_evolutionary_2009}. The other cost function is defined by Genç in \cite{genc_bearing-only_2008}, namely equidistant line segments. Basically, assuming constant velocity and course, the sum of absolute deviations of the line segments between the observation points is calculated as the cost. In \cite{tokta_target_2017}, the cost function is defined as a hybrid function of these two and showed its effectiveness with Monte Carlo simulations.

In this work, these two cost functions used in literature, namely bearing differences which is used in the nonlinear estimation algorithms and equidistant lines which is used in the Cartesian and pseudolinear filters which use state space representation, for the TMA problem are investigated. When equidistant line segments is used, which means estimation in state space (or linearized), polar to Cartesian conversion is needed. However, this is a nonlinear transformation and leads to bias for the transformed measurements and this violates the basic assumption that the measurement errors have Gaussian densities with zero mean. Therefore, the cost function based on equidistant line segments will often give a sub-optimal solution. We analyze the target-ownship geometries that lead to this bias, measure the degree of Gaussianity of that geometry, and show that the bias causes unreasonable estimation errors for specific geometries when equidistant line segments is used.

The rest of the paper is structured as follows. In section \ref{TMA} we present the TMA problem formulation and the parameters that affect the target observability are stated and the cost functions are presented. In section \ref{sims}, for some target-ownship geometries for Bearings-Only Target Motion Analysis (BO-TMA) problem is simulated using these cost functions and they are compared against each other in terms of position estimation errors. In section \ref{bias}, the results of the previous section are analyzed and it is shown, by measuring the degree of Gaussianity of the geometry, that the results of the previous section is actually caused by bias introduced by the coordinate transformation. In section \ref{conc} the conclusions are given.

\section{TMA Problem Definition}
\label{TMA}

\begin{figure}[t!]
\begin{center}
\includegraphics[width=9.0cm, trim={0 1cm 0 1cm}]{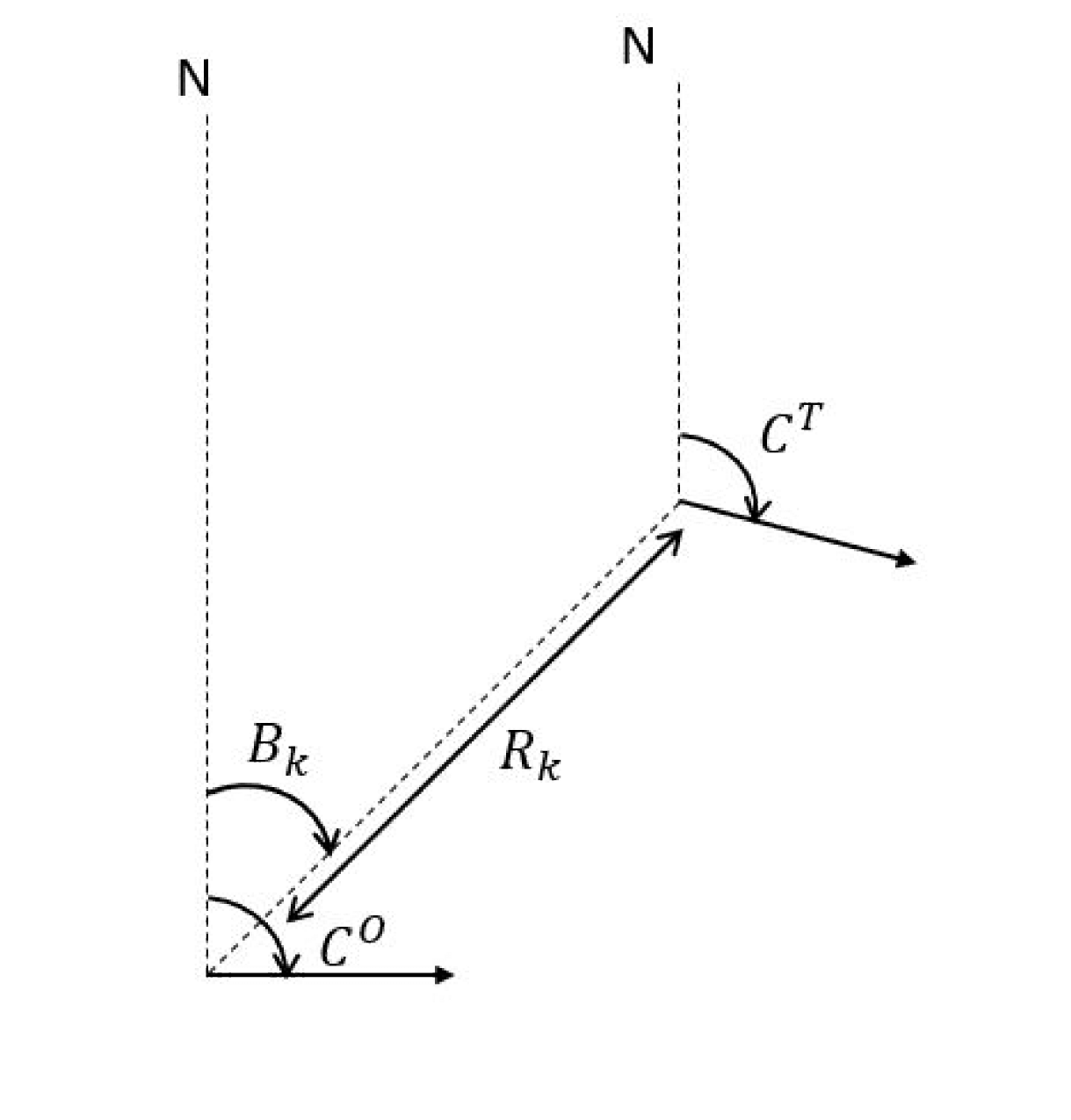}
\caption{Target-Observer geometry.}
\label{fig1}
\end{center}\vs{-8mm}
\end{figure}

The bearing-only target tracking problem is to estimate the target trajectory using only the  bearing measurements, which is the angle between the target and the moving observer (ownship) from the North, in clockwise. Bearing angles are measured in \textit{$T_s$} intervals. Target-Observer geometry is shown in Figure \ref{fig1}. In Figure \ref{fig1}, \textit{$B_k$} is the measured bearing angle, \textit{$R$} is the range between the target, where subscript \textit{k} is the sample index. The observer, \textit{$C_k^O$} and \textit{$C^T$} are the course of the observer and the target, respectively. The bearing measurements are subject to white Gaussian noise. The TMA problem is, using only the bearing measurements which are collected at every \textit{$T_s$} seconds, to estimate the target position or the target motion parameters, i.e. target range, course and speed.

The position of the target at each measurement time is calculated as follows.

\begin{equation}
\label{eq1}
\begin{bmatrix}
\hat{X}_{0}^T \\
\hat{Y}_{0}^T \\
\end{bmatrix}
= 
\begin{bmatrix}
X_0^O \\
Y_0^O \\
\end{bmatrix}
+
\begin{bmatrix}
\sin(B_0) \\
\cos(B_0) \\
\end{bmatrix}
R_0.
\end{equation}
where \textit{$X_0$} and \textit{$Y_0$} denote the initial position of the target in Cartesian coordinates, \textit{$B_0$} is the initial bearing measurement and \textit{$R_0$} is the initial range. After calculating the initial position, velocity and position of the target at any time \textit{$t=kT_s$} can be calculated from the target parameters as in (\ref{eq2}).

\begin{equation}
\label{eq2}
\begin{bmatrix}
\hat{V}_{x,k}^T \\ 
\hat{V}_{y,k}^T \\
\end{bmatrix}
= 
S \begin{bmatrix}
sin(C^T)\\
cos(C^T)
\end{bmatrix},
\begin{bmatrix}
\hat{X}_k^T \\
\hat{Y}_k^T 
\end{bmatrix}
=
\begin{bmatrix}
\hat{X}_0^T + T_s \hat{V}_{x,k}^T  \\
\hat{Y}_0^T + T_s \hat{V}_{y,k}^T
\end{bmatrix}.
\end{equation}
where \textit{S} denotes the speed of the target and ( $\hat{}$ ) above the variable indicates estimated value of the variable. The bearing angle $\theta$, between the target and the ownship, at sample \textit{k}, measured from the North (which is the positive Y axis on Cartesian coordinate system) is given as

\begin{equation}
\label{eq3}
\theta_k = arctan \Big(\frac{Y_k^T - Y_k^O}{X_k^T - X_k^O}\Big).
\end{equation}
Hence, if the target parameters \textit{$R_k$}, \textit{S} and \textit{$C_k^T$} is known, the target trajectory can be found. As suggested in \cite{song_observability_1996,le_cadre_discrete-time_1997}, in order for the target to be observable, the ownship must outmaneuver the target which means ownship must have a higher order derivative than that of the target.

\begin{figure}[t!]
\begin{center}
\includegraphics[width=14.0cm, trim={2cm 9cm 2cm 9cm}]{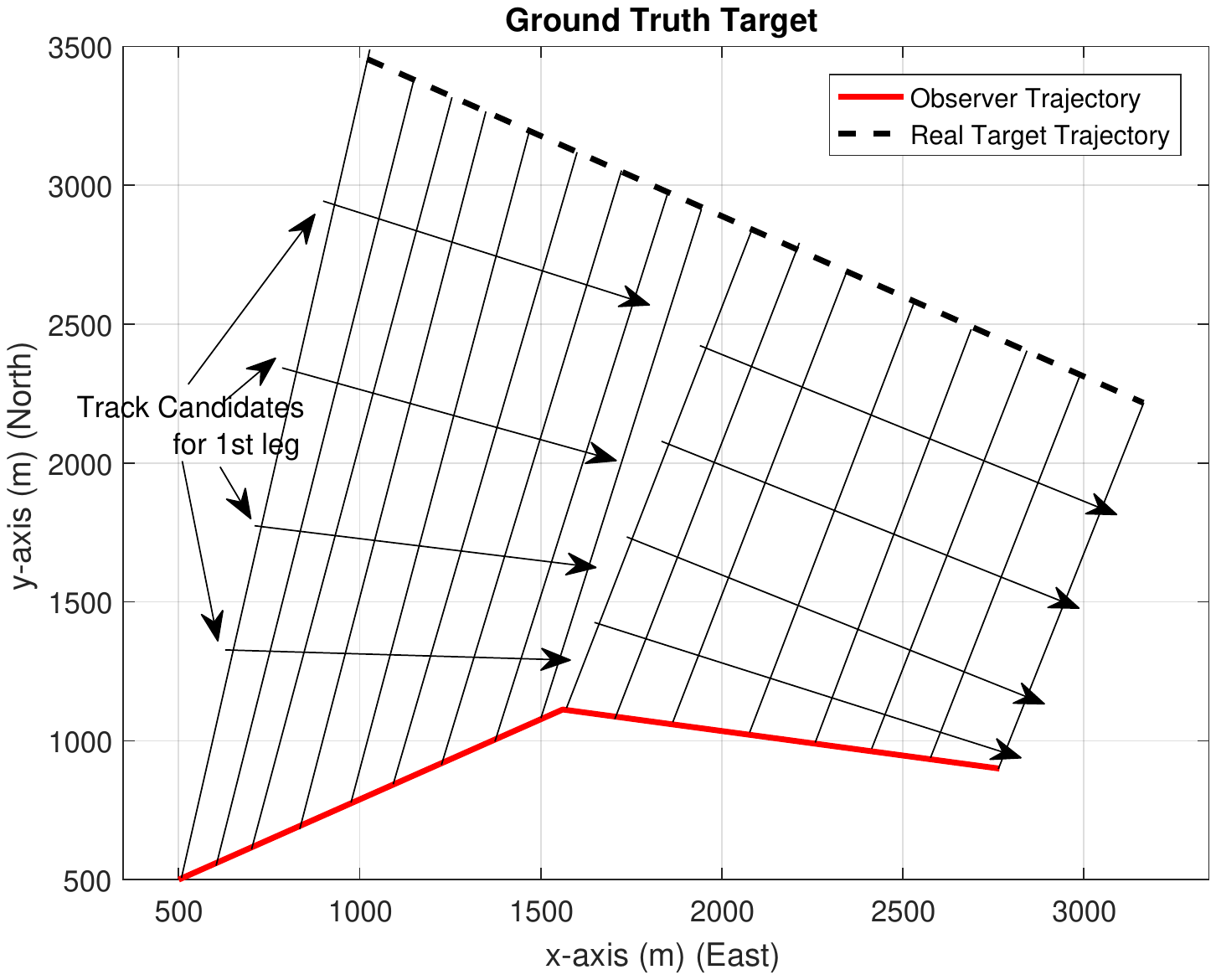}
\caption{Target-Observer geometry with target track candidates.}
\label{fig2}
\end{center}\vs{-8mm}
\end{figure}

In Figure \ref{fig2}, Bearing Only TMA problem geometry is given. The vertical lines show bearing lines between the target and the ownship. Also, possible solutions for the target trajectory are shown in Figure \ref{fig2}, which are track candidates. In the first leg of the ownship, there are infinitely many candidates for the target trajectory. In other words, there are infinitely many choices for the target trajectory that would lead to the equal distances, at different ranges. When the ownship starts to collect bearing measurements in the second leg, the solution becomes unique as only one trajectory lead to equal distances between the intersection of the bearing lines.

In addition to target observability, the degree of observability is an important issue. The degree of observability may come up in terms of the difference between two consecutive bearing measurements, in other words, bearing rate. High bearing rate corresponds to a highly observable target. The bearing rate is strongly related to the range and speed of the target. Therefore, close targets with low speed have lower bearing rate, while the targets that are far away (with low or high speed) have larger bearing rate. Another parameter that affect the bearing rate is the course of the target (or together with speed, velocity of the target). If the course of the target is perpendicular to the line of sight (or the bearing line), observability increases.

\subsection{Overview of the Cost Functions}

To obtain a solution for the target trajectory in BO-TMA, an error criterion is needed. The available error criteria (or the cost functions) in literature are point differences when Cartesian coordinates are used and  bearing differences when nonlinear methods are used. In this section, estimation error or optimization criteria for the BO-TMA are presented.

The first cost function is equidistant line segments \cite{genc_new_2010}. Assuming bearing measurements are collected at constant time intervals, for a target which moves at a constant speed, the bearing lines divide the line in the direction of the target course at equal distances. In other words, the target has the same displacement at every measuring time. The cost function is defined as the deviation from the mean distance of line segments as follows
\begin{equation}
\label{eq4}
D_{mean} = \frac{ \sum\limits_{k=1}^{N-1}\left | P_k - P_{k+1} \right | }{N-1}.
\end{equation}
where $P_k=[X_k^T, Y_k^T]'$ is the point of intersection of the $k^{th}$ bearing line and candidate target trajectory and $N$ is the total number of bearings. Equation (\ref{eq4}) gives the mean distance between consecutive intersection points. However, the mean distance of line segments may become larger or smaller for some target-ownship geometries, i.e. when the target and ownship velocities are different. Therefore, the lengths of the line segments must be normalized w.r.t. the mean as in

\begin{equation}
D_{normalized,k} = \frac{D_k-D_{mean}}{D_{mean}}.
\end{equation}
where $D_k = P_k - P_{k+1}$. After obtaining normalized distances, the cost function to be minimized can be calculated as in
\begin{equation}
\label{eq6}
Cost = \sum\limits_{k=1}^{N-1}\left| D_{normalized,k}\right|.
\end{equation}
Notice that the cost function (\ref{eq6}) only uses two target parameters, namely, $R_0$ and $C^T$. The target speed can be calculated by dividing the total distance by the total time elapsed.

The cost function in (\ref{eq6}) inherently assumes Gaussian distributed distances. However, when the bearing measurements are noisy, the distances between bearing lines are not normally distributed, due to the bias introduced by the polar to Cartesian transformation.

The other error criterion that appear in literature is the bearing differences which is used in nonlinear filters, i.e. extended Kalman filter \cite{spingarn_passive_1987,xu_single_2004}. This is calculated as the square of difference between the measured bearing and the estimated bearings, at time $t_k$ as in
\begin{equation}
\label{eq7}
\Delta \theta_k = (\theta_k- \hat{\theta}_k)^2.
\end{equation}
where $\theta_k$ is the measured bearing and $\hat{\theta}_k$ is the estimated bearing at time $t_k$. Summing the squared differences in (\ref{eq7}) for the $N$ measurements and taking the square root of the sum gives the cost associated with the estimates of $\hat{\theta}_k, k=1,...,N$.
\begin{equation}
\label{eq8}
Cost = \sqrt[]{\sum\limits_{k = 1}^{N}\Delta\theta_k} .
\end{equation}
The cost functions given in (\ref{eq6}) and (\ref{eq8}) are formulated here for the batch processing techniques. However, their equivalents for recursive filters also exist.


\section{Simulations}
\label{sims}

In this section, we show that, for some target-ownship geometries, the position estimation error increases when (\ref{eq6}) is used, therefore we suggest that it is more appropriate, to use the cost function in (\ref{eq8}).

To show this effect, we have simulated BO-TMA problem for three parameters in order to estimate the target trajectory. These parameters are initial range of the target $R_0$, initial bearing $B_0$ and target course $C^T$. Though there are other parameters (i.e. speed of the target and ownship, ownship course) that affect the quality of the estimation, only $R_0$, $B_0$ and $C^T$ are considered to keep the analysis relatively simple.

The initial bearing and ownship courses were kept constant. As for the range parameter, three different values $5000m$, $15000m$, $25000m$, were chosen as "short range", "mid range" and "long range", respectively. Initial bearing is kept constant, $B_0= 45^{\circ}$ in all simulations The target parameters used in simulations are given in the Table \ref{tab1}.


The simulations were conducted to estimate the target parameters $R_0$, $C^T$ and $S$. In order to eliminate algorithm dependency and guarantee a solution, a search space was defined around the true solution and the optimum solution was searched for, with respect to the cost functions defined in the previous section, using the brute force approach, thus the method is algorithm-free. For the range parameter, the search space was defined between [$R_0-1000m$, $R_0+1000m$] meters with $10$ meter-steps. The target course was searched between $C^T\pm 2^{\circ}$ , with $0.1^{\circ}$ steps. As for the speed, the parameter was searched between [1, 20] $m/s$. For each scenario, 20 Monte Carlo simulations were run and average of these runs were assessed as results. For both cost functions, two errors, one is the parameter error which is the difference between the true parameter vector and estimated parameter vector, and the other is the position estimation error which is the root mean squared (RMS) error of the target trajectory, which is generated from the estimated parameter vector, against the true target trajectory, were calculated. In all scenarios, the target speed was kept constant, $5 m/s$. The bearing noise is Gaussian and its standard deviation is also constant $1^{\circ}$, in all simulations.

\begin{figure}[t!]
\begin{center}
\includegraphics[width=15.0cm, trim={0cm 1cm 0cm 1cm}]{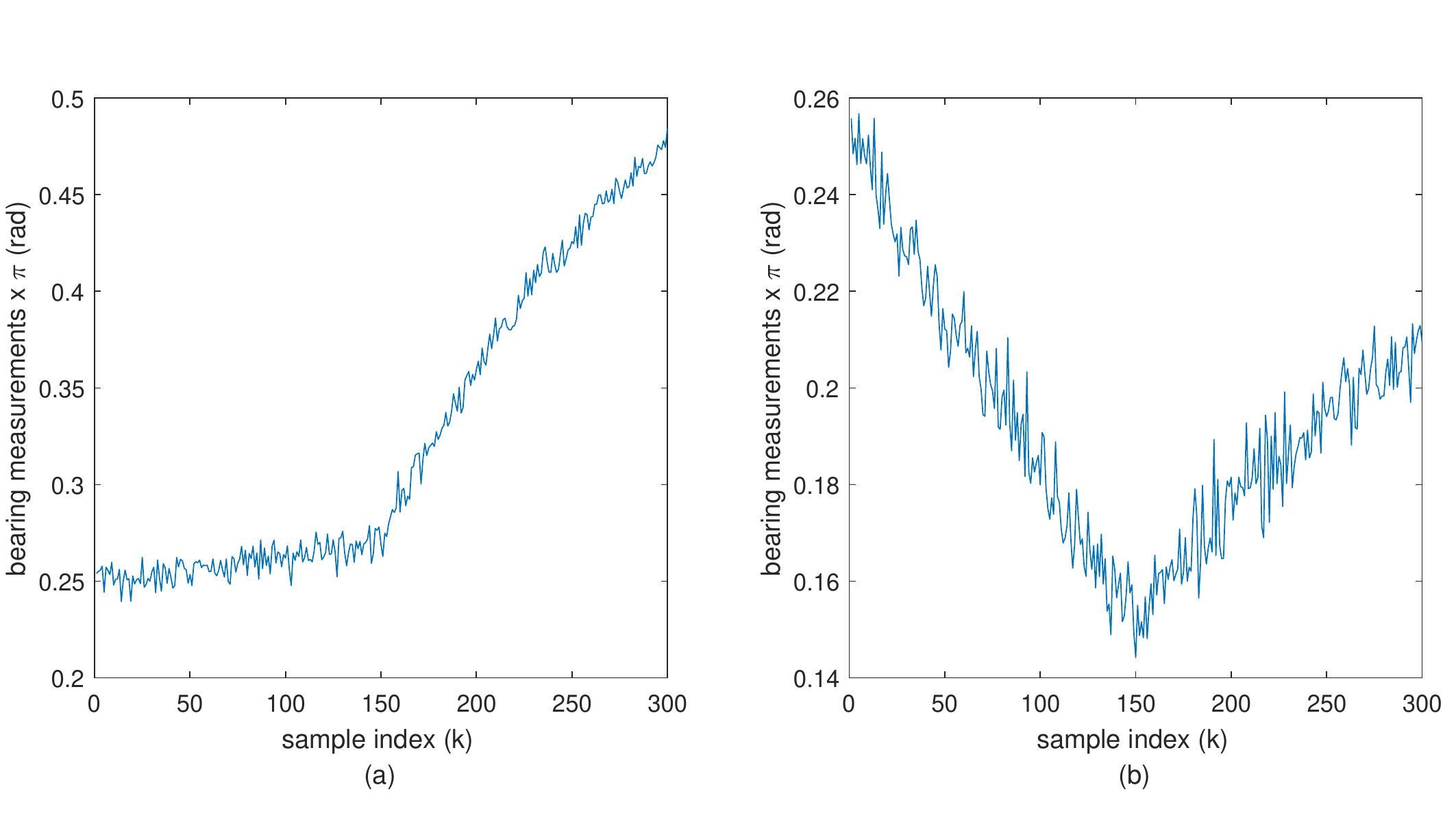}
\caption{Scenario-1: Short range, (a) bearing rate is small, (b) bearing rate is large.}
\label{fig3}
\end{center}\vs{-6mm}
\end{figure}

In scenario 1, for a constant range and ownship course, the effect of target course on the estimation performance was assessed. Ownship course $C^O = 100^\circ$ is selected. Bearing measurements are plotted in Figure \ref{fig3} (a) for $C^T = 170^{\circ}$ and (b) for $C^T = 30^{\circ}$. As can be seen from the Figure \ref{fig3}, bearing change is smaller for $C^T = 30^{\circ}$ and larger for $C^T = 170^{\circ}$.

\begin{table}[h]
\begin{center}\vs{-4mm}
\caption{Scenario 1 results}\vs{1mm}
\label{tab2}
\begin{tabular}{|l|c|c|c|c|}
\hline
\raisebox{0ex}[0.5cm][0cm]{}& \multicolumn{2}{|c|}{\textbf{Bearing Difference (\ref{eq8})}} & \multicolumn{2}{|c|}{\textbf{Equidistant (\ref{eq6})}} \\ \hline
\raisebox{0ex}[0.5cm][0cm] {$R_0 = 5000 m$, $B_0 = 45^{\circ}$} & $C^T = 30^{\circ}$ & $C^T = 170^{\circ}$ & $C^T = 30^{\circ}$ & $C^T = 170^{\circ}$		\\ \hline
\raisebox{0ex}[0.5cm][0cm]{\textbf{Parameter Error}} & [0 0.1 0] & [0 0 0] & [640 2 0.1] & [200 1 0.09]     \\ \hline
\raisebox{0ex}[0.5cm][0cm]{\textbf{RMS Error}} & 75$m$ & 0$m$ & 709$m$ & 170$m$ \\ \hline
\end{tabular}
\end{center}\vs{-6mm}
\end{table}

In Table \ref{tab2}, parameter error was given as the difference of the obtained parameter vector from the true parameter vector, in the form of $[R_0, C^T, S]$. RMS error was calculated after the target trajectory had been generated. For a constant range, when $C^T = 30^{\circ}$, the bearing change is smaller. When $C^T = 170^{\circ}$, the bearing change is larger than for $C^T = 30^{\circ}$. Therefore, the performance of the cost function (\ref{eq6}) is relatively good. However, when the cost function (\ref{eq8}) is used, there is no significant difference in estimation error. Results are unaffected. Overall, bearing differences (\ref{eq8}) gives better estimation performance than the equidistant line segments (\ref{eq6}). Besides, the Table \ref{tab2} shows that when initial bearing and target course are close, the performance decreases. However, while this has no significant affect on the cost function (\ref{eq8}), the performance of the cost function (\ref{eq6}) is much worse, even in the close range. So, the cost function (\ref{eq6}) is more sensitive to the difference between the initial bearing and the target course.

In scenario 2, the effect of the target range on the estimation performance was assessed. For different range values and for a constant course, the simulations were repeated. Obtained results were given in Table \ref{tab3}.

\begin{table}[h!]
\begin{center}\vs{-4mm}
\caption{Scenario 2 results}\vs{1mm}
\label{tab3}
\begin{tabular}{|l|c|c|c|c|}
\hline
\raisebox{0ex}[0.5cm][0cm]{}& \multicolumn{2}{|c|}{\textbf{Bearing Difference} (\ref{eq8})} & \multicolumn{2}{|c|}{\textbf{Equidistant} (\ref{eq6})} \\ \hline
\raisebox{0ex}[0.5cm][0cm] {$C^T = 30^{\circ}$} & \textbf{Parameter Error} & \textbf{RMS Error} & \textbf{Parameter Error} & \textbf{RMS Error}\\ \hline
\raisebox{0ex}[0.5cm][0cm]{\textbf{R}$_0 = 5000 m$}  & [0 0 0]  & $0m$ & [583 2 0.01] & $658m$ \\ \hline
\raisebox{0ex}[0.5cm][0cm]{\textbf{R}$_0 = 15000 m$} & [100 0 0] & $100m$ & [447 2 0.02]  & $359m$ \\ \hline
\raisebox{0ex}[0.5cm][0cm]{\textbf{R}$_0 = 25000 m$} & [100 1 0]  & $134m$ & [210 1 0.2]   & $273m$ \\ \hline
\end{tabular}
\end{center}\vs{-6mm}
\end{table}
When the range increases, obviously, the estimation performance is expected to decrease, since the noise on the bearing measurements will have greater effect. However, for (\ref{eq6}), as it is seen in Table \ref{tab3}, the error gets smaller as the range increases. This result is contradicts with expected result, for the cost function (\ref{eq6}).


From these results, it can be said that the cost function (\ref{eq8}) is better than (\ref{eq6}). However, (\ref{eq6}) is almost as good as (\ref{eq8}) for determining the target speed, regardless of the quality of the initial range estimation and the estimation error does not seem to increase much when (\ref{eq6}) is used.

In the next section, we will further analyze these geometries, in terms of bias introduced by the coordinate transformation and Gaussianity of the error parameter to be estimated for the equidistant cost function. We will show that the obtained results are related to transformation bias.

\section{Bias Analysis}
\label{bias}
In this section, we further analyze the geometries that we have simulated in the previous section and interpret the results from a different view. Aidala stated in \cite{aidala_utilization_1983} that the use of Cartesian filters generate biased estimates because of the noisy measurements. This is due to the nonlinear nature of the polar to Cartesian transformation of noisy bearing measurements. When noisy bearings are transformed into Cartesian, the transformed variables are no longer Gaussian (unless the bearing noise is uniform).

\subsection{Measure of Gaussianity}
Considering this statement, we have simulated certain geometries and determined how biased the transformed measurements (or decision variables) are. In order to do this, we need to measure the amount of bias introduced by coordinate transformation. The bearing measurements (random variables) are Gaussian and when the coordinate transformation is applied, obtained random variables deviate from the Gauss. We quantify this deviation and identify the relation between obtained results and biasedness of the geometry.

Amongst some choices, i.e. Kullback-Liebler divergence (relative entropy) \cite{kullback1951}, Hellinger distance \cite{hellinger_neue_1909} which need a reference distribution, kurtosis measure is used to measure how Gaussian is probability distribution of the transformed random variable. Kurtosis is related to the tails of the distribution and for a univariate normal distribution its kurtosis $\kappa = 3$, and it can be calculated from (\ref{eq9}).

\begin{equation}
\label{eq9}
Kurtosis[X] = \kappa = E\bigg[\bigg(\frac{X-\mu}{\sigma}\bigg)^4\bigg] = \frac{\mu_4}{\sigma^4}.
\end{equation}
Where $\mu$ is the expected value, $\mu_4$ is the $4^{th}$ moment and $\sigma$ is the standard deviation  of $X$. The argument of expectation operator is standardization of random variable $X$. Therefore, kurtosis, actually, measures "outliers" which are data outside of one standard deviation around the mean. Higher kurtosis is the result of infrequent and extreme deviations. So, using kurtosis, we could measure how Gaussian a probability distribution of a random variable is, in other words whether the distribution has heavy tails or not. When the distribution of a random variable deviates from the Gauss, its kurtosis deviates from the reference value of a Gaussian random variable's.

\subsection{Bias simulations}
To simulate transformation bias, the BO-TMA problem was set up for two points, for both target and ownship. For a given constant range $R_0$, and constant target and ownship speeds, we generated a target-ownship geometry for some possible target and ownship course combinations. Briefly, for a given target-ownship geometry, we generate bearing measurements for two points and from the noisy measurements, by applying coordinate transformation, target positions were generated and target displacement is measured. The magnitude of this displacement, actually, corresponds to equidistant line segment. This is repeated for $10^5$ Monte Carlo runs and consequently, an empirical distribution for equidistant line segments is obtained. Ideally, the obtained distribution must be constant (because of the constant speed assumption). By measuring kurtosis of this distribution, we obtain a quantitative information about how Gaussian the distribution is, indirectly, how biased are the transformed data.

\begin{figure}[t!]
\begin{center}
\includegraphics[width=18.0cm, trim={0cm 1cm 0cm 1cm}]{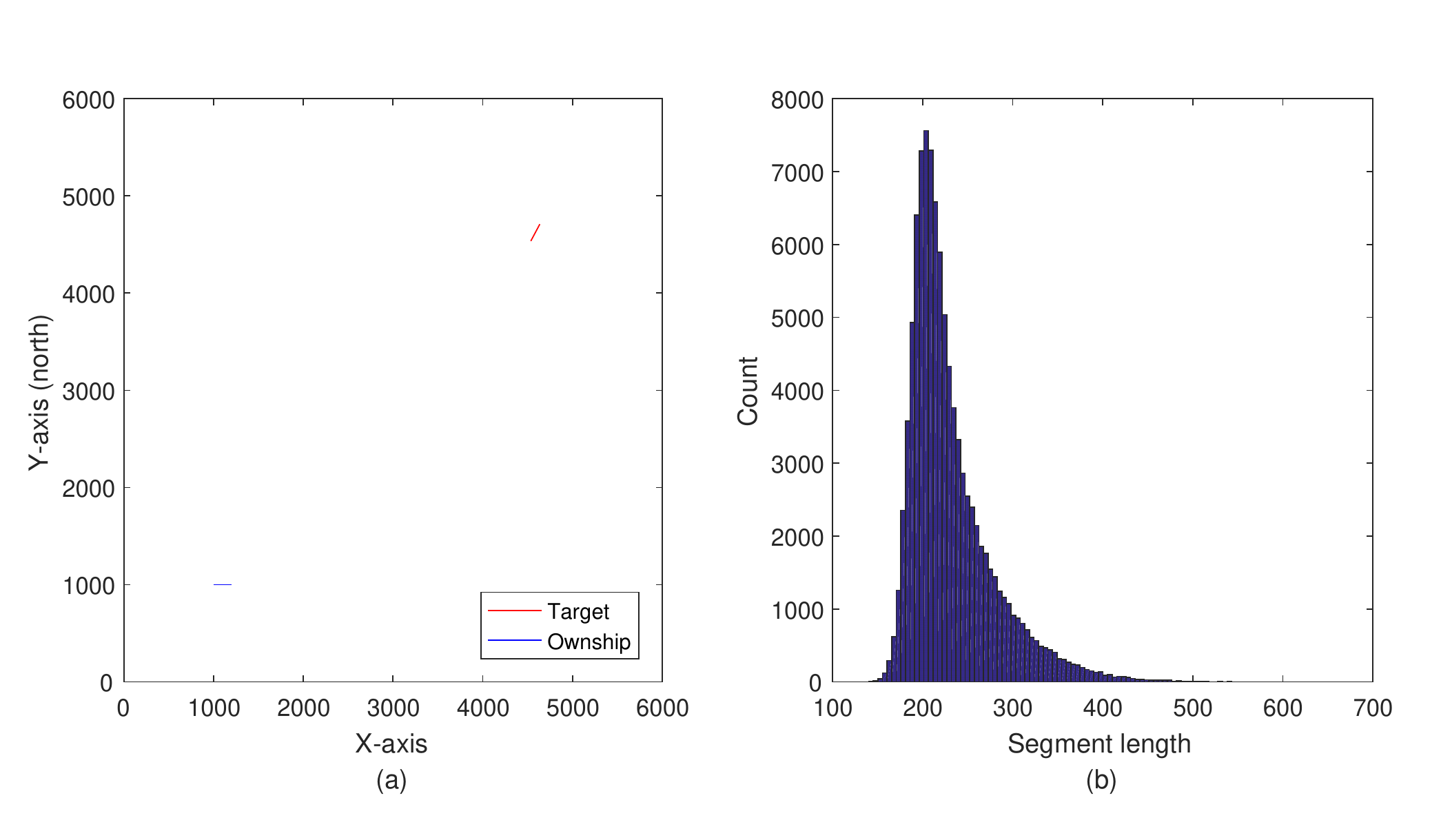}
\caption{$R_0 = 5000m$, $C^O = 90^{\circ}$ and $C^T = 30^{\circ}$ (a) target-ownship geometry, (b) equidistant segment distribution.}
\label{fig4}
\end{center}\vs{-6mm}
\end{figure}
\begin{figure}[t!]
\begin{center}
\includegraphics[width=18.0cm, trim={0cm 1cm 0cm 1cm}]{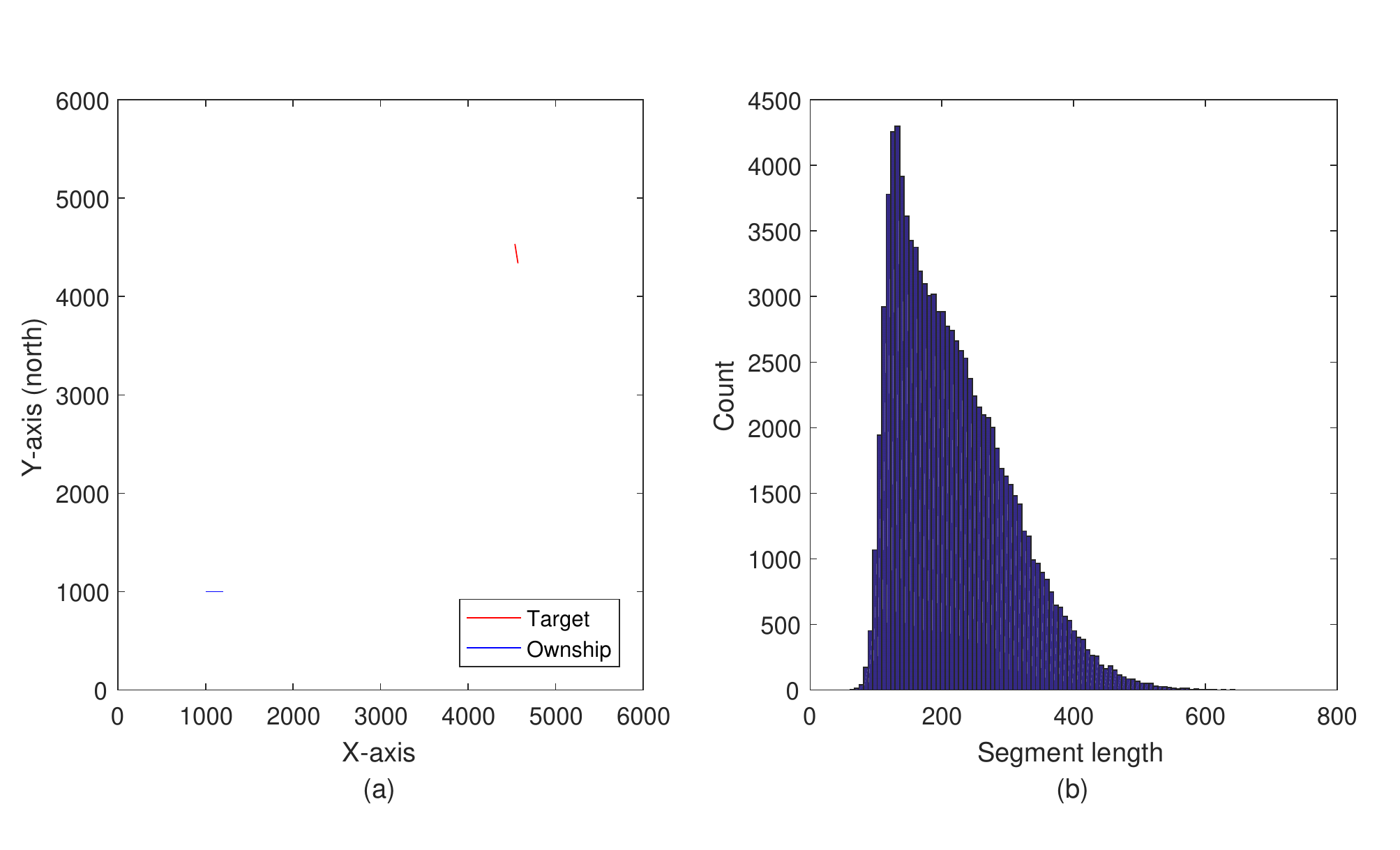}
\caption{$R_0 = 5000m$, $C^O = 90^{\circ}$ and $C^T = 170^{\circ}$ (a) target-ownship geometry, (b) equidistant segment distribution.}
\label{fig5}
\end{center}\vs{-6mm}
\end{figure}

In Figure \ref{fig4}, (a) the target-ownship geometry and (b) estimated target displacement distribution is shown. This is scenario 1 from previous section for $C^T = 30^{\circ}$. As can be seen from Figure \ref{fig4}-(b), the distribution is, obviously, biased. Measured kurtosis for this distribution is $\kappa = 6.9$. So, this tells that the obtained distribution is far from Gaussian. In Figure \ref{fig5}, $C^T = 170^{\circ}$ case of scenario 1 is simulated. The measured kurtosis for this case is $\kappa = 3.3$. There is a significant difference between two geometries. When $C^T = 170^{\circ}$, the distribution of the equidistant line segments is closer to a Gaussian, so the geometry matches with the inherent Gaussian assumption of the cost function (\ref{eq6}) and provides an accurate estimation for the target parameters.

This result tells us that the reason behind the improvement in the RMS error (Table \ref{tab2}, equidistant column) is the effect of reduced bias of this geometry. $C^T = 30^{\circ}$ case is more biased (less Gaussian) than the $C^T = 170^{\circ}$ case, with respect to the kurtosis measure. Therefore, as the difference between the target course and the bearing line increases, the estimation performance also increases. The effect of bias introduced by the coordinate transformation is less significant for this target-ownship geometry.

Likewise, this is what happens when the range increases (scenario 2). As the range increases, kurtosis of the distribution of the segment becomes closer to the kurtosis of a Gaussian, which means the bias is less significant. The measured kurtosis values for different ranges are given in Table \ref{tab4}.

\begin{figure}
\begin{center}
\includegraphics[width=12.0cm, trim={0cm 1cm 0cm 1cm}]{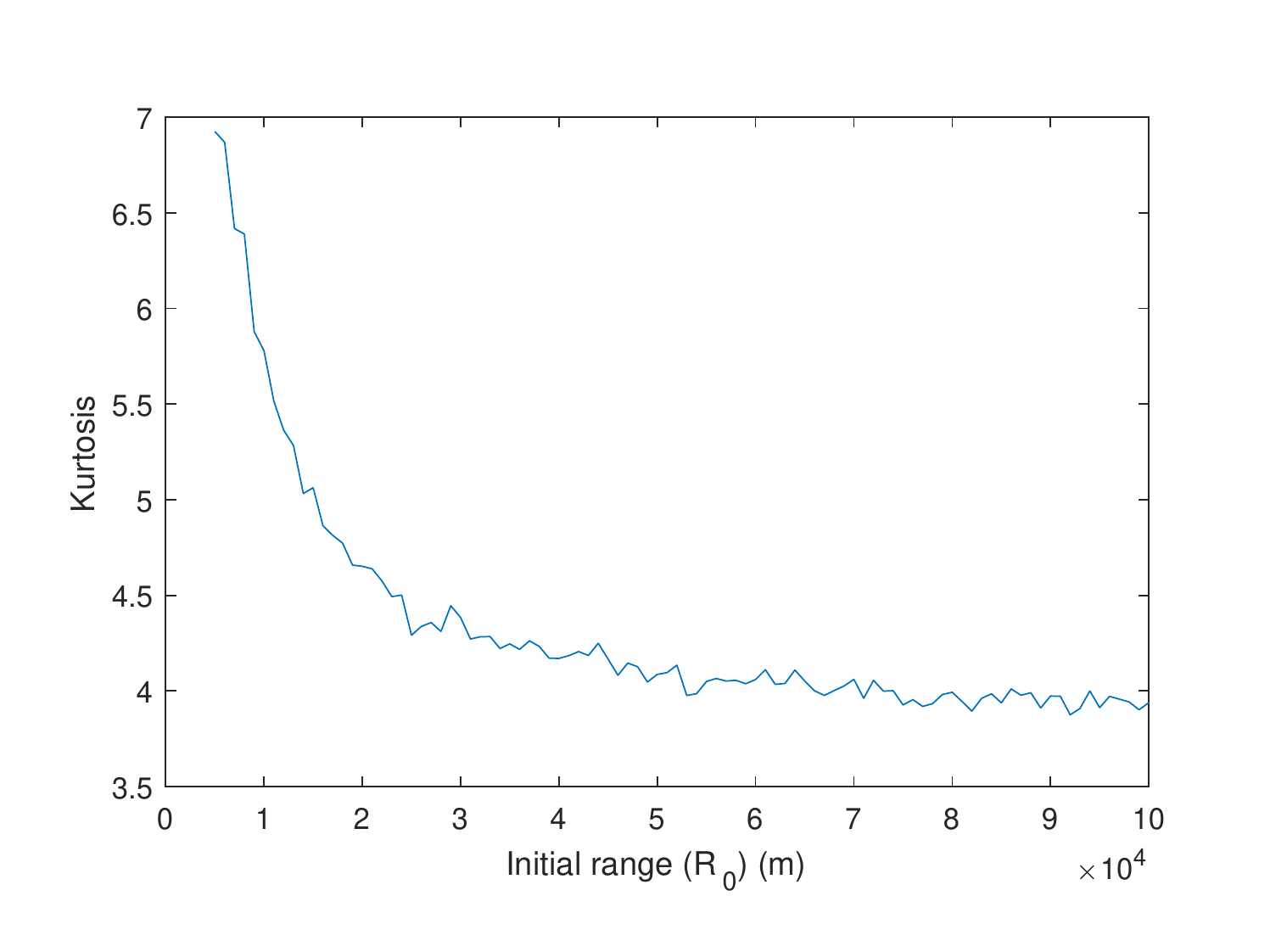}
\caption{Kurtosis value against range from $R_0 = 5000m$ to $R_0 = 10^5m$.}
\label{fig6}
\end{center}\vs{-6mm}
\end{figure}

\begin{table}[h!]
\begin{center}\vs{-4mm}
\caption{Measured kurtosis for different range values.}\vs{1mm}
\label{tab4}
\begin{tabular}{|l|c|c|c|c|}
\hline
\raisebox{0ex}[0.5cm][0cm]{$C^T=30^{\circ}$} & $R_0 = 5000m$ & $R_0 = 25000m$ & $R_0 = 50000m$ & $R_0 = 100000m$ \\ \hline
\raisebox{0ex}[0.5cm][0cm]{\textbf{kurtosis ($\kappa$)}} & 6.9 & 4.4 & 4.09 & 3.95 \\ \hline
\end{tabular}
\end{center}\vs{-6mm}
\end{table}
The change in Kurtosis value is most significant when range changes $R_0 = 5000m$ to $R_0=25000m$. After that, it remains almost constant. To see the decrease in kurtosis against the range, we have simulated from $R_0 = 5000m$ to $R_0 = 100000m$ with $1000m$ steps, for $C^T = 30^{\circ}$ and $C^O = 90^{\circ}$ case and calculated kurtosis of the distribution, obtained from Monte Carlo runs. In Figure \ref{fig6} obtained kurtosis values for each range value is plotted.

As it can be seen from the Figure \ref{fig6}, as the range increases kurtosis value decreases until it settles about $4$. In other words, the segment distribution becomes more like a Gaussian as the range increases. Therefore, the performance of equidistant line segments gets better for increasing range values.

The simulations conducted in this section verifies the results of previous section. Analysis of this section shows that the results obtained in previous section is actually caused by transformation bias. These results can be interpreted as, for certain target-ownship geometries, the effect of this bias is reduced. So, for different target geometries different cost functions can be preferred. For example, when the target range (and bearing rate) is small, the cost function (\ref{eq8}) performs better estimation accuracy. While the range increases, the estimation performance of the equidistant lines cost function (\ref{eq6}) approaches that of (\ref{eq8}) and can be preferred for mid range scenarios.

\section{Conclusions}
\label{conc}
In this paper, we studied the transformation bias present in BO-TMA problem. Main contribution of this paper is that we specify the target-ownship geometries that reduces the effect of transformation bias for a cost function and effect of this bias on the estimation error. We measured the bias, in terms of degree of Gaussianity of a distribution, using kurtosis statistics. We showed that to reduce the effect of coordinate transformation bias on the estimation performance, the target course must be different than the bearing line. In addition, we investigated the effect of range on this bias and showed that as the range increases, the effect of bias decreases until it settles to a certain value, in terms of kurtosis. The estimation performance increases as the decision variable of the cost function (especially for equidistant line segments) becomes more like a Gaussian.

\section*{Acknowledgment}
.

\bibliographystyle{unsrt}
\bibliography{tma_bibtex}

\end{document}